\documentclass[12pt]{article}

\usepackage{mathrsfs}
\usepackage{amssymb}
\usepackage{amsmath}
\usepackage{mathtools}
\usepackage{ascmac}
\usepackage{amsthm}
\usepackage{algorithm}
\usepackage{algorithmic}
\usepackage{enumitem}
\usepackage{bm}
\usepackage{graphicx}
\usepackage{enumerate}
\usepackage{natbib}
\usepackage{xcolor}
\usepackage{url} 

\newcommand{\blind}{0}

\theoremstyle{definition}

\newcommand{\mX}{\mathcal{X}}

\renewcommand{\hat}{\widehat}

\newcommand{\argmin}{\operatornamewithlimits{argmin}}

\begin{document}

\def\spacingset#1{\renewcommand{\baselinestretch}%
{#1}\small\normalsize} \spacingset{1}


\if0\blind
{
  \title{\bf Trend Filtering for Functional Data}
  \author{Tomoya Wakayama\thanks{
    Corresponding author, Email: tom-w9@g.ecc.u-tokyo.ac.jp}\hspace{.2cm}\\
    Graduate School of Economics, The University of Tokyo\\
    and \\
    Shonosuke Sugasawa \\
    Center for Spatial Information Science, The University of Tokyo}
  \maketitle
} \fi

\if1\blind
{
  \bigskip
  \bigskip
  \bigskip
  \begin{center}
    {\LARGE\bf Locally Adaptive Smoothing for\\ Functional Data}
\end{center}
  \medskip
} \fi

\bigskip
\begin{abstract}
Despite increasing accessibility to function data, effective methods for flexibly estimating underlying functional trend are still scarce. We thereby develop functional version of trend filtering for estimating trend of functional data indexed by time or on general graph by extending the conventional trend filtering, a powerful nonparametric trend estimation technique, for scalar data. We formulate the new trend filtering by introducing penalty terms based on $L_2$-norm of the differences of adjacent trend functions.
We develop an efficient iteration algorithm for optimizing the objective function obtained by orthonormal basis expansion. 
Furthermore, we introduce additional penalty terms to eliminate redundant basis functions, which leads to automatic adaptation of the number of basis functions. The tuning parameter in the proposed method is selected via cross validation.
We demonstrate the proposed method through simulation studies and applications to real world datasets.
\end{abstract}

\noindent%
\vfill

\newpage
\spacingset{1.5} 

\section{Introduction}
Due to advances in measurement devices and data storage resources, it is nowadays possible to observe functions as realizations of random experiments and thus functional data analysis (FDA) has expanded rapidly in recent decades. Functional versions for many branches of statistics have been provided, for example, in \cite{ramsay2004functional}, \cite{kokoszka2017introduction} and \cite{horvath2012inference}.

The conventional techniques of FDA for independent functional data have been recently extended to dependent situations (both time series and spatial cases). 
In fact, for functional time series data, standard stationary models for multivariate data have been extended \citep[e.g.][]{besse2000autoregressive,klepsch2017innovations,klepsch2017prediction,hormann2013functional,gao2019high,hormann2015dynamic} and theoretical properties have also been widely investigated \citep[e.g.][]{bosq2000linear,aue2017estimating,spangenberg2013strictly,aue2017functional,kuhnert2020functional,cerovecki2019functional}. 
On the other hand, effective estimations of functional trend under non-stationary situations are not well developed despite their importance in real applications. 
\cite{van2018locally} addressed a framework for locally stationary functional times series, but its flexibility for trend estimation is still limited. 
Regarding spatial functional data, while spatial interpolation methods under spatial stationary have been developed \citep[e.g.][]{giraldo2011ordinary,nerini2010cokriging}, there are some attempts to estimate non-stationary spatial trend determined by some external covariates \citep[e.g.][]{caballero2013universal,menafoglio2013universal,menafoglio2016universal}.
However, many flexible estimation methods for spatially varying trend are not considered.

Although many useful tools are available in FDA, the flexibility of existing methods may be limited; that is, handling abrupt changes in a trend is challenging.
Hence the need for locally adaptive smoothing methods arises.
For univariate time series, trend filtering \citep{kim2009ell_1,tibshirani2014adaptive} is recognized as a powerful tool for locally adaptive trend estimation. Additionally, \cite{wang2016trend} extended trend filtering to spatial data, which enables us to estimate spatial trend with abrupt changes revealed.

In this work, we provide an effective local smoothing method for functional time series data by extending trend filtering for scalar data.
Combining $L_2$-loss and $L_2$-norm penalty terms for differences of adjacent functions, we successfully define the objective function for functional trend filtering. To solve the optimization problem, we expand the functional data via orthonormal basis functions and transform the objective function. 
We find that this transformed objective function is a mixture of the fused lasso \citep{tibshirani2005sparsity} and the grouped lasso \citep{yuan2006model,lounici2011oracle,tibshirani1996regression}. This is rather different from the case of scalar, where only fused lasso-like penalties are considered. We then develop an iterative algorithm based on the idea of ADMM \citep{boyd2011distributed,ramdas2016fast}, in which each updating procedure can be easily carried out. 
Furthermore, to satisfy a demand for selecting the optimal number of basis functions, we additionally construct an trend estimator. This also contributes to the denoising of the observed functional data.
We also extend functional trend filtering from time series data to data on a graph and analyze spatial functional data. 
As for the tuning parameter selection, we simply suggest using cross validation, which is fairly feasible owing to the efficient optimization algorithm.

The remainder of the paper is organized as follows. 
Section~\ref{sec:TF} offers a brief review of trend filtering and its periphery, which is deeply related to our work. 
In Section~\ref{sec:method}, we present the methods, functional trend filtering, for both functional time series and spatial data, and describe the algorithm to carry out the proposed method. Also we discuss the selection of the number of basis functions.
In Section~\ref{sec:sim}, we compare the proposed method with some existing approaches through simulation studies.
In Section~\ref{sec:app}, we apply the proposed method to functional time series (fertility rates as a function of age in each year) and functional spatial data (the number of confirmed COVID-19 cases as a function of day in Japanese prefectures).
Finally we conclude with a discussion in Section~\ref{sec:conc}


\section{Review of trend filtering for scalar data}
\label{sec:TF}
Before describing the proposed methods for functional data, we briefly review trend filtering known as a powerful tool for locally adaptive smoothing for scalar time series data.
Let $y_1,\ldots,y_T$ be a sequence of observations, and we are interested in denoising the observations to estimate the underlying trend denoted by $\beta=(\beta_1,...,\beta_T)^{\top}$. 
The $k$th order trend filtering \citep{kim2009ell_1,tibshirani2014adaptive} is defined as the minimizer of the following objective function:
\begin{align}\label{TF-scalar}
\frac{1}{2} \sum_{t=1}^T (y_t-\beta_t)^2  + \lambda\sum_{t=1}^{T-k-1}\lvert \Delta_t^{(k)}\beta\rvert,
\end{align}
where $\lambda\geq 0$ is a tuning parameter which controls the trade-off between the fit to the observed data and smoothing the trend estimation.
Here $\Delta_t^{(k)}$ is the $t$th row vector of the $k$th order discrete difference operator matrix $\Delta^{(k)}$ defined as  
\begin{align*}
\Delta^{(k)}\coloneqq \begin{cases}
    D^{(0)} & {\rm for} \,\,k=0 ,\\
    D^{(k)}\Delta^{(k-1)} & {\rm for} \,\,k\geq 1,
 \end{cases}
\end{align*}
where $D^{(k)}$ is the following $ (T-k-1) \times (T-k)$ matrix:
\begin{align*}
    D^{(k)} &\coloneqq\left(
        \begin{array}{cccccc}
          1 & -1& 0&\ldots & 0 &0\\
          0 & 1 & -1&\ldots & 0 &0\\
          \vdots & \vdots&\vdots & \ddots & \vdots&\vdots \\
          0 & 0&0 & \ldots & 1&-1
        \end{array}
        \right)
\end{align*}
For example, (\ref{TF-scalar}) with $k=0$ is the same form of fused lasso \citep{tibshirani2005sparsity} and the penalty makes many differences to zero exactly and leave others nonzero values, leading to piece-wise constant estimation of $\beta$. 
In general, sparsity of $\beta$ under $k$th order discrete difference operator matrix suggests that the estimated components have a specific $k$th order piece-wise polynomial structure \citep{tibshirani2014adaptive}. While trend filtering is locally adaptive estimator defined by a regularization problem with nonsmooth penalty, it is still computationally efficient owing to its convexity.

Trend filtering is also applicable to spatial data \citep{wang2016trend}.
Let $V=\{1,\dots,n\}$ be a set of sample index, which can be regard as vertex of graph $G=(V,E)$.
Here $E=\{e_1,\dots,e_m\}$ is a set of undirected edges according to the spatial adjacent structure, where $e_h\in V\times V$ for $h=1,\ldots,m$ and $m$ is the total number of adjacency relationships. 
For example, $e_h=(i,j)$ means that $i$th and $j$th locations are adjacent. 
Let $\Delta^{(0)}\in \{1,0,-1\}^{m\times n}$ be the oriented incidence matrix of the graph $G$, that is, $\Delta^{(0)}_{hi}=1$, $\Delta^{(0)}_{hj}=-1$ and the other elements in the $h$th row vector of $\Delta^{(0)}$ is $0$ if $e_h=(i,j)$.
We then define 
\begin{align*}
\Delta^{(k+1)}\coloneqq
\begin{cases}
(\Delta^{(0)})^{\top} \Delta^{(k)} & {\rm for \,\,even}\,\, k ,\\
\Delta^{(0)}\Delta^{(k)} & {\rm for\,\, odd}\,\, k.
\end{cases}
\end{align*}

This $\Delta^{(k)}$ is hereinafter referred to as the $k$th order graph difference operator matrix. 
The $k$th order spatial trend filtering estimate is obtained as the minimizer of the following function: 
\begin{align*}
\frac{1}{2} \sum_{i=1}^n (y_i-\beta_i)^2 +\lambda\| \Delta^{(k)}\beta\|_1,
\end{align*}
where $\lambda$ is a tuning parameter. We remark it is also a form of fused lasso and accordingly it can be solved by basic convex optimization algorithms. \cite{wang2016trend} discusses the computational aspect in detail. 
Notably, the penalty term encourages sparsity in graph differences in trend, which yields a piece-wise polynomial nature of the estimator as the original trend filtering (\ref{TF-scalar}).

\section{Functional trend filtering}
\label{sec:method}
We will develop the method discussed above into functional data, that is, find a trend among functions.

\subsection{\textit{Settings and objective function}}
Let $(\Omega,\mathcal{A},P)$ be an arbitrary probability space. The space $L^2(\mathcal{X})$ is defined as the set of all real valued square integrable functions on a compact set $\mX\subset\mathbb{R}$. It is a Hilbert space with norm $\|f\|_{L^2}=(\int_{\mX} f^2(x)dx)^{1/2}$, which is induced  by the inner product $\langle f,g\rangle=\int_{\mX}f(x)g(x)dx$ for $f,g\in L^2(\mX)$. We consider a serially indexed collection $\{Y_t(\cdot):t=1,...,T\}$ of random functions defined on the same probability space: $Y_t:\Omega\rightarrow L^2(\mathcal{X})$ is a measurable map. $\{y_t(\cdot):t=1,...,T\}$ denote a set of observed functions. The trend of $\beta_t=E[Y_t]$ as a function of $t$ is of interest. To estimate $\beta=(\beta_1,\dots,\beta_T)^{\top}\in(L^2(\mathcal{X}))^T$, we propose the $k$th order $functional\, trend \, filtering$ defined as
\begin{align}
\label{ell1}
\hat{\beta}^{\rm TF}
&=\argmin_{\beta} 
 \left[\frac{1}{2} \sum_{t=1}^T \int_{\mathcal{X}}\left\{y_t(x)-\beta_t(x))\right\}^2 dx + \lambda\sum_{t=1}^{T-k-1}\left[\int_{\mathcal{X}} \left\{\Delta_t^{(k)}\beta(x)\right\}^2dx\right]^{1/2}\right]\\
&=\argmin_{\beta} 
\left\{\frac{1}{2} \sum_{t=1}^T \|y_t-\beta_t\|_{L^2}^2 + \lambda\sum_{t=1}^{T-k-1}\|\Delta_t^{(k)}\beta\|_{L^2}\right\}\nonumber,
\end{align}
where $\lambda>0$ is a tuning parameter and $\Delta_t^{(k)}$ is the $t$th row vector of the $k$th order difference operator matrix $\Delta^{(k)}$. 
For instance, if $k=0$, the penalty is $\sum_{t=1}^{T-1} \|\beta_t-\beta_{t+1}\|_{L^2}$, which can be regarded as the functional version of the group fused lasso \citep{alaiz2013group}. Henceforth, we have to solve the functional version of the group fused lasso with general order $k$, desiring that some elements of $\{\Delta_t^{(k)}\hat{\beta}^{\rm TF} : t=1,...,T-k-1\}$ are set to zero.

Since the observations and the true functions are infinite dimension and difficult to handle, we first prepare $L$ orthonormal basis functions $\{\phi_{\ell}:\ell=1,\dots,L\}$ on $L^2(\mX)$, which satisfy
\[
  \langle\phi_{\ell} ,\phi_{\ell'}\rangle = \begin{cases}
    1 & \mathrm{if} \,\,\,\ell=\ell' ,\\
    0 & \mathrm{otherwise}.
  \end{cases}
\] 
Then, using the approximate expansions 
$y_t(x)\approx\sum_{\ell=1}^L z_{t\ell}\phi_{\ell}(x)$ and $\beta_t(x)\approx \sum_{\ell=1}^L b_{t\ell}\phi_{\ell}(x)$ with $z_{t\ell}=\langle y_t,\phi_{\ell}\rangle$ and $b_{t\ell}=\langle \beta_t,\phi_{\ell}\rangle$ for all $t$ and $\ell$, we reduce problem (\ref{ell1}) to minimization of the objective function
\begin{equation*}
    \frac12\sum_{t=1}^T\sum_{\ell=1}^L (z_{t\ell}-b_{t\ell})^2 + \lambda\sum_{t=1}^{T-k-1}\left[\sum_{\ell=1}^L(\Delta_t^{(k)}\bm{b}_\ell)^2\right]^{1/2},
\end{equation*} 
with respect to $\bm{b}_\ell=(b_{1\ell},\ldots,b_{T\ell})^{\top}\in\mathbb{R}^T$.
Define $B=(\bm{b}_1,...,\bm{b}_L)$ and $\bm{z}_{\ell}=(z_{1\ell},\ldots,z_{T\ell})^{\top}$.
The above objective function can be further rewritten as
\begin{equation}\label{obj}
\frac12\sum_{\ell=1}^L\|\bm{z}_{\ell}-\bm{b}_{\ell}\|_2^2+\lambda \sum_{t=1}^{T-k-1}\|\Delta_t^{(k)}B\|_2,
\end{equation}
where $\|\cdot\|_2$ denotes the $\ell_2$ norm. Focusing on the latter half, we notice it is a mixture of $k$th order fused lasso type penalty and grouped lasso type penalty.
In the case of scalar \citep{kim2009ell_1,tibshirani2014adaptive}, it is enough to consider fused lasso-like penalty, but in the case of functional data, mixture of group lasso and fused lasso-like penalty is necessary.

\cite{alaiz2013group} extended group lasso to a fused setting and addressed the solution, but the order of fusion is limited to $k=0$.
Both $\|\cdot\|_2^2$ and $\|\cdot\|_2$ are convex functions and $\|\bm{z}_{\ell}-\bm{b}_{\ell}\|_2^2$ is differentiable with respect to $\bm{b}_{\ell}$, but $\|\Delta_t^{(k)}B\|_2$ is not separable, namely, it cannot be represented as sum of the univariate convex function of each $b_{t\ell}$. Accordingly, solving this problem is not straightforward extension of scalar version and requires ingenuity.

\subsection{\textit{Optimization}}\label{sec:opt}

For notational simplicity, we use $\Delta$ instead of $\Delta^{(k)}$ in what follows. 
To optimize (\ref{obj}), we first introduce two unit vectors, $\bm{e}_t^{a}\in \mathbb{R}^{T-k-1}$ and $\bm{e}_t^{b}\in \mathbb{R}^{L}$, whose only $t$th and $\ell$th elements are $1$, respectively, and the other elements are 0.

Since $\bm{e}_{\ell}^b\Delta_tB=\bm{e}_t^a \Delta\bm{b}_{\ell}$, we rewrite the optimization of (\ref{obj}) with respect to $\{\bm{b}_\ell\}$ as the following constraint optimization problem: 
\begin{align*}
&\argmin_{\{\bm{a}_t\},\{\bm{b}_{\ell}\}} \left\{\frac12\sum_{\ell=1}^L\|\bm{z}_{\ell}-\bm{b}_{\ell}\|_2^2+\lambda \sum_{t=1}^{T-k-1}\|\bm{a}_t\|_2\right\} \\
& \qquad\mathrm{subject \ to} \ \ \ \ \bm{e}_{\ell}^b\bm{a}_t=\bm{e}_t^a \Delta \bm{b}_{\ell} \ \ \ \ \ (\ell=1,...,L \ {\rm and} \ t=1,...,T-k-1).
\end{align*}
Note that the objective function is similar to one for alternating direction method of multipliers (ADMM) algorithm \citep{boyd2011distributed,ramdas2016fast}, which breaks the problem into smaller pieces that are easier to deal with. 
We then define an augmented Lagrangian function 
\begin{align}
    &L_{\rho}\,(\{\bm{a}_t\},\{\bm{b}_{\ell}\},\{\bm{u}_t\}):= \frac12\sum_{\ell=1}^L\|\bm{z}_{\ell}-\bm{b}_{\ell}\|_2^2+\lambda \sum_{t=1}^{T-k-1}\|\bm{a}_t\|_2  \nonumber \\ 
    \,\,\,\,\,\,&+\sum_{t=1}^{T-k-1}\sum_{\ell=1}^L u_{t\ell}(\bm{e}_t^a\Delta\bm{b}_{\ell}-\bm{e}_{\ell}^b\bm{a}_t)+\frac{\rho}{2}\sum_{t=1}^{T-k-1}\sum_{\ell=1}^L(\bm{e}_t^a\Delta\bm{b}_{\ell}-\bm{e}_{\ell}^b\bm{a}_t)^2,
\label{augmented}\end{align}
where $\{u_{t\ell}\}_{t,\ell}$ is Lagrange multipliers and $\rho>0$ controls the influence of the violation of equality constraint. Since there is no apparent closed form solution for $\bm{b}_{\ell}$ minimizing the objective function $L_{\rho}$, we develop an iterative algorithm outlined in Algorithm 1, where the derivation is deferred to Appendix. The convergence of Algorithm 1 is empirically confirmed.
From the existing theory of ADMMirically confirmed. algorithm, the choice of $\rho$ is related only to the speed of convergence of the algorithm without affecting the final estimates \citep[e.g.][]{boyd2011distributed,fukushima1992application,he2000alternating}. 
In our implementation, we simply set $\rho=0.1$.

Using the coefficients $\hat{\bm{b}}^{\rm TF}_{\ell}$ computed by the procedure, we obtain the function
\begin{align*}
    \hat{\beta}^{\rm TF}_t(x)= \sum_{\ell=1}^L \hat{b}^{\rm TF}_{t\ell}\phi_{\ell}(x)  
\end{align*} for $t=1,...,T$, which is the estimator of the trend.

\begin{algorithm}                      
\caption{(Functional trend filtering)}         
\label{alg1}
\renewcommand{\algorithmicrequire}{\textbf{Input:}}
\renewcommand{\algorithmicensure}{\textbf{Initialization:}}
\begin{algorithmic}
\vspace{2mm}
\REQUIRE $\bm{z}_1,\ldots,\bm{z}_L$ (coefficient vector), $\Delta$ (difference operator), $\varepsilon_0, \varepsilon_1$ (tolerance level)\\
\vspace{2mm}
\ENSURE{$b_{t\ell}^{(0)},\,\,a_{t\ell}^{(0)},\,\, u_{t\ell}^{(0)}$}
\vspace{2.5mm}
\WHILE{$ \sum_{\ell=1}^L\|\bm{b}_{\ell}^{(v+1)}-\bm{b}_{\ell}^{(v)}\|_{2}\geq \varepsilon_0 $ }
\vspace{2mm}
\STATE $<update \,\,\bm{b}_{\ell}>$
\FOR{$\ell=1,...,L$}
\vspace{0.5mm}
\STATE  $\bm{b}_{\ell}^{(v+1)}  \leftarrow
    (I+\rho \Delta^{\top}\Delta)^{-1}\{\bm{z}_{\ell}  -  \sum_{t}u_{t\ell}^{(v)}(\bm{e}_t^a\Delta)^{\top}  +  \rho\sum_{t}(\bm{e}_t^a\Delta)^{\top}\bm{e}_{\ell}^b\bm{a}_t^{(v)}\}$
\vspace{0.5mm}
\ENDFOR
\vspace{2mm}
\STATE $<update \,\,\bm{a}_t>$
\vspace{0.5mm}
\FOR{$t=1,...,T-k-1$}

\STATE  $\hat{\bm{w}}_0=\bm{a}_t^{(v)}\in\mathbb{R}^L,\,s_0\leftarrow1,\,j\leftarrow0,\,\eta\leftarrow1$
\vspace{0.8mm}
\WHILE{$\|\bm{w}_{j+1}-\bm{w}_j\|_2\geq \varepsilon_1 $}
\vspace{0.8mm}
\STATE $\bm{w}_{j+1}\leftarrow S_{\lambda}\{(1-\rho)\hat{\bm{w}}_j+\rho\sum_{\ell}\bm{e}_{\ell}^b{}^{\top}(\bm{e}_t^a\Delta\bm{b}_{\ell}^{(v+1)}+\rho^{-1}u_{t\ell}^{(v)})\}$ 
\vspace{0.5mm}
\STATE \hspace{2cm}  with $S_{\lambda}(\bm{s}):=\max(0, 1-\lambda/\|\bm{s}\|_2)\bm{s}$
\vspace{0.5mm}
\STATE $s_{j+1}\leftarrow (1+\sqrt{1+4s_j^2})/2$
\vspace{0.5mm}
\STATE $\hat{\bm{w}}_{j+1}\leftarrow \bm{w}_{j+1}+(s_j-1)(\bm{w}_{j+1}-\bm{w}_{j})/s_{j+1}$
\ENDWHILE
\vspace{0.5mm}
\STATE $\bm{a}_t^{(v+1)}\leftarrow \bm{w}_j$
\ENDFOR
\vspace{2mm}
\STATE $<update \,\,u_{t\ell}>$
\FOR{$t=1,...,T-k-1$ and $\ell=1,...,L$}
\vspace{0.5mm}
\STATE $u_{t\ell}^{(v+1)} \leftarrow u_{t\ell}^{(v)}+\rho(\bm{e}_t^a\Delta\bm{b}_{\ell}-\bm{e}_{\ell}^b\bm{a}_t)$
\vspace{0.5mm}
\ENDFOR
\ENDWHILE
\end{algorithmic}
\end{algorithm}

As an alternative smoother, we construct a simplified version of the trend estimation
\begin{align}
    &\hat{\beta}^{\rm HP}=\argmin_{\beta} \frac{1}{2} \sum_{t=1}^T \int_{\mathcal{X}}(y_t(x)-\beta_t(x))^2 dx + \lambda\sum_{t=1}^{T-k-1}\int_{\mathcal{X}} \left\{\Delta_t\beta(x)\right\}^2dx\label{ell2}.
\end{align}
The difference between (\ref{ell2}) and (\ref{ell1}) is the penalty. Specifically, the penalty in (\ref{ell2}) is the squared value of the $L^2$-norm used in (\ref{ell1}).  
Since the estimator defined in (\ref{ell2}) can be regarded as an extension of Hodrick-Prescott (HP) filter \citep{10.2307/2953682}, we refer to this method as functional HP filter.
Although the use of squared norm penalty does not produce sparsity in the differences, the method is easy to implement.
In fact, by expanding (\ref{ell2}) via orthonormal functions, we have the following approximation of the objective function: 
\begin{align}
     \frac12\sum_{\ell=1}^L\|\bm{z}_{\ell}-\bm{b}_{\ell}\|_2^2+\lambda \sum_{t=1}^{T-k-1}\sum_{\ell=1}^L(\Delta_t\bm{b}_{\ell})^2\label{ell2red},
\end{align}
which yields the closed form solution given by 
\begin{align*}
    \hat{\bm{b}}^{\rm HP}_{\ell}=\Big(I_T+2\sum_{t=1}^{T-k-1}\lambda\Delta_t^{\top}\Delta_t\Big)^{-1}\bm{z}_{\ell}, \ \ \ \ 
\end{align*}$\ell=1,\ldots,L.$
In some situations, it works better than functional trend filtering. (Details are given in Section 4.)

Finally, we discuss the choice of tuning parameter $\lambda$. In practice, the value of $\lambda$ used for filtering is determined by $K$-fold cross validation, where we divide the dataset into $K$ subsets by extracting every $K$th function. As the estimate of $\bm{b}_t$ in a validation dataset, we take the midpoint of $\bm{b}_{t-1}$ and $\bm{b}_{t+1}$ after smoothing.

\subsection{\textit{Extension to functional data on graph}}

Let $G=(V,E)$ be the graph with vertices $V=\{1,\dots,n\}$ and undirected edges $E=\{e_1,\dots,e_m\}$, representing spatial adjacent structure.
Let $\{Y_i(\cdot): 1,\ldots,n\}$ be random functions on the vertices, which take values in the space $L^2(\mX)$ on a compact set $\mX\subset\mathbb{R}$.
Suppose that $E[Y_i(x)]=\beta_i(x)$ for $i=1,\cdots,n$ and we are interested in the estimation of $\beta_i(x)$.
Let $\Delta^{(k)}$ be $k$th order graph difference operator matrix defined in Section \ref{sec:TF}. 
We propose the $k$th order $functional\, trend \, filtering\, on \,graph$ to estimate $\beta=(\beta_1,\dots,\beta_{n})^{\top}$ by
\begin{align}
    &\hat{\beta}^{\rm TF}=\argmin_{\beta}      \frac{1}{2} \sum_{i=1}^n \int_{\mathcal{X}}(y_i(x)-\beta_i(x))^2 dx \nonumber\\
    &\qquad \qquad \quad+\lambda \sum_{p=1}^q\,\left[\int_{\mathcal{X}}\{\Delta^{(k)}_{p}\beta(x)\}^2dx\right]^{1/2}
    \label{tfsp}
\end{align}
where $q=n$ if $k$ is odd, $q=m$ otherwise. The penalty quantifies how much $\beta$ vary locally in the sense of $k$th order graph differences.
We prepare $L$ orthonormal basis functions $\phi_1(x),\phi_2(x),\ldots,\phi_L(x)$ and approximate $y_i(x)\approx\sum_{\ell=1}^L z_{i\ell}\phi_{\ell}(x)$ and $\beta_i(x)\approx \sum_{\ell=1}^L b_{i\ell}\phi_{\ell}(x)$ with $z_{i\ell}=\langle y_i,\phi_{\ell}\rangle$ and $\,b_{i\ell}=\langle \beta_i,\phi_{\ell}\rangle$ for all $i$ and $\ell$. 
This is an extension of \citep{wang2016trend} to functional data, but the optimization is far more complicated, as we showed in time series.

In the following discussion in this section, we write $\Delta$ for $\Delta^{(k)}_p$.
Define two standard unit vectors $\bm{e}_p^{a}\in \mathbb{R}^{q}$ and $\bm{e}_{\ell}^{b}\in \mathbb{R}^{L}$, whose only $t$th and $\ell$th elements are $1$, respectively, and the other elements are 0, and $\bm{b}_\ell=(b_{1\ell},\ldots,b_{n\ell})^{\top}\in\mathbb{R}^n$.
Following the same logic as the previous section, we regard the problem to find (\ref{tfsp}) as a problem to get $\{\bm{b}_{\ell}\}$ minimizing an augmented Lagrangian, for a parameter $\rho>0$,
\begin{align*}\small
    &L_{\rho}\,(\{\bm{a}_p\},\{\bm{b}_{\ell}\},\{\bm{u}_p\}):= \frac12\sum_{\ell=1}^L\|\bm{z}_{\ell}-\bm{b}_{\ell}\|_2^2
    +\lambda \sum_{p=1}^{q}\|\bm{a}_p\|_2 \\
    &\quad +\sum_{p=1}^q\sum_{\ell=1}^L u_{p\ell}(\bm{e}_p^a\Delta\bm{b}_{\ell}-\bm{e}_{\ell}^b\bm{a}_p)+\frac{\rho}{2}\sum_{p=1}^q\sum_{\ell=1}^L(\bm{e}_p^a\Delta\bm{b}_{\ell}-\bm{e}_{\ell}^b\bm{a}_p)^2.
\end{align*}
To solve this problem, we can again utilize Algorithm 1.
Using the acquired coefficients $\hat{\bm{b}}^{\rm TF}_{\ell}$ computed by the procedure, we obtain the function
\begin{align*}
    \hat{\beta}^{\rm TF}_i(x)= \sum_{\ell=1}^L \hat{b}^{\rm TF}_{i\ell}\phi_{\ell}(x),
\end{align*}
for $i=1,...,n$. This is the estimator of the proposed method.

As an alternative smoothing method, we also propose an estimator:
\begin{align}
  &\hat{\beta}^{\rm HP}=\argmin_{\beta} \frac{1}{2} \sum_{t=1}^n \int_{\mathcal{X}}(y_t(x)-\beta_t(x))^2 dx  +\lambda \sum_{p=1}^q\,\int_{\mathcal{X}}\{\Delta_{p}\beta(x)\}^2dx.
\end{align}
It corresponds to (\ref{ell2}) in time series setting, or, Laplacian regularization \citep{smola2003kernels} for univariate data.
By treatment with the same approximation as the former section, we convert the problem into the optimization problem with objective function:
\begin{align*}
  &\frac12\sum_{i=1}^n\sum_{\ell=1}^L (z_{i\ell}-b_{i\ell})^2 + \lambda\sum_{p=1}^{q}\sum_{\ell=1}^L(\Delta_p b_\ell)^2.
\end{align*}
For $\ell=1,...,L$, the closed form solution is given by
\begin{align*}
  \hat{\bm{b}}^{\rm HP}_{\ell}=\Big(I_n+2\sum_{p=1}^{q}\lambda\Delta_p ^{\top}\Delta_p\Big)^{-1}\bm{z}_{\ell}.
\end{align*}
Consequently, we obtain the estimator $\hat{\beta}^{\rm HP}_i(x)= \sum_{\ell=1}^L \hat{b}^{\rm HP}_{i\ell}\phi_{\ell}(x)$ for $i=1,...,n$.

\subsection{\textit{Selection of the number of basis via additional regularization}}
\label{sec:NoB}

\begin{algorithm}                      
\caption{(Sparse functional trend filtering)}         
\label{alg1}
\renewcommand{\algorithmicrequire}{\textbf{Input:}}
\renewcommand{\algorithmicensure}{\textbf{Initialization:}}
\begin{algorithmic}
\vspace{2mm}
\REQUIRE $\bm{z}_1,\ldots,\bm{z}_L$ (coefficient vector), $\Delta$ (difference operator), $\varepsilon_0, \varepsilon_1, \varepsilon_2$ (tolerance level)\\
\vspace{2mm}
\ENSURE{$b_{t\ell}^{(0)},\,\,a_{t\ell}^{(0)},\,\, u_{t\ell}^{(0)}$}
\vspace{2.5mm}
\WHILE{$ \sum_{\ell=1}^L\|\bm{b}_{\ell}^{(v+1)}-\bm{b}_{\ell}^{(v)}\|_{2}\geq \varepsilon_0 $ }
\vspace{2mm}
\STATE $<update \,\,\bm{b}_{\ell}>$
\FOR{$l=1,...,L$}
\STATE  $\hat{\bm{w}}_0=\bm{b}_{\ell}^{(v)}\in\mathbb{R}^L,\,s_0\leftarrow1,\,j\leftarrow0,\,\eta\leftarrow1$
\vspace{0.8mm}
\WHILE{$\|\bm{w}_{j+1}-\bm{w}_j\|_2\geq \varepsilon_2 $}
\vspace{0.8mm}
\STATE $\bm{w}_{j+1}\leftarrow S_{\sigma^{-1}_{sp}}\{(I+\rho \Delta^{\top}\Delta)\hat{\bm{w}}_{\ell} - \bm{z}_{\ell}  +  \sum_{t}u_{t\ell}(\bm{e}_t^a\Delta)^{\top}-\rho\sum_{t}(\bm{e}_t^a\Delta)^{\top}\bm{e}_{\ell}^b\bm{a}_t\}$ 
\vspace{0.5mm}
\STATE \hspace{2cm}  with $S_{\lambda}(\bm{s}):=\max(0, 1-\lambda/\|\bm{s}\|_2)\bm{s}$
\vspace{0.5mm}
\STATE $s_{j+1}\leftarrow (1+\sqrt{1+4s_j^2})/2$
\vspace{0.5mm}
\STATE $\hat{\bm{w}}_{j+1}\leftarrow \bm{w}_{j+1}+(s_j-1)(\bm{w}_{j+1}-\bm{w}_{j})/s_{j+1}$
\ENDWHILE
\vspace{0.5mm}
\STATE $\bm{b}_{\ell}^{(v+1)}\leftarrow \bm{w}_j$
\ENDFOR

\vspace{2mm}
\STATE $<update \,\,\bm{a}_t>$
Same as Algorithm1.
\vspace{2mm}
\STATE $<update \,\,u_{t\ell}>$
Same as Algorithm1.
\ENDWHILE
\end{algorithmic}
\end{algorithm}

The methods introduced so far are established by orthonormal basis expansion. What we need to be careful about is the necessity of choosing the number of basis functions in practice since the optimal number of basis functions depends on the complexity of the function.
Here, we solve this challenge by constructing the following estimator:
\begin{align}\label{eq:sftf}
    &\hat{\beta}^{\rm SFTF}=\argmin_{\beta} \frac12\sum_{t=1}^T\sum_{\ell=1}^L (z_{t\ell}-b_{t\ell})^2
    + \lambda\sum_{t=1}^{T-k-1}\left[\sum_{\ell=1}^L(\Delta_t^{(k)}\bm{b}_\ell)^2\right]^{1/2}
    + \psi \sum_{\ell=1}^L \omega_{\ell}\|\bm{b}_\ell\|_2,
\end{align}where $\lambda$ and $\psi$ are tuning parameters, and $\omega_\ell$ is a fixed weight for the $\ell$th coefficients.
When we use the principal components as basis functions, we set $\omega_\ell$ as the inverse values of the proportion of variance. The crux of this method is the last $L$ terms. These are in the form of a group lasso, where the coefficients of each basis function are one group.
This makes all the coefficients of the unnecessary basis zero and only the necessary part remains, thus allowing the selection of the number of basis. In practice, we select many basis functions beforehand and regard survivors as the essential basis functions. Also reducing unnecessary principal components promotes smoothing with respect to $x$-direction of the function while the middle terms of (\ref{eq:sftf}) make estimator smooth with respect to $t$-direction. In what follows, this method is referred as $sparse\,functional\,trend\,filtering$.

For the algorithm of this method, it is sufficient to modify the way $\{\bm{b}_{\ell}\}$ is updated in Algorithm 1. See Appendix 2 for details. We also select the additional tuning parameter $\psi$ by $K$-fold cross validation.

\section{Simulation Studies}
\label{sec:sim}
In the previous chapters, we develop the two methods. An overview of the simulation is presented in Section \ref{sec:sim:pro}. 
In Section \ref{sec:sim:ftf}, to give a fair comparison of functional trend filtering and other methods, we fix the same number of basis functions for all methods.
In Section \ref{sec:sim:sftf}, we implement sparse functional trend filtering and another method and compare their performance.

\subsection{\textit{Procedure}}
\label{sec:sim:pro}
\begin{figure}[tb]
  \begin{center}
  \includegraphics[width=16cm]{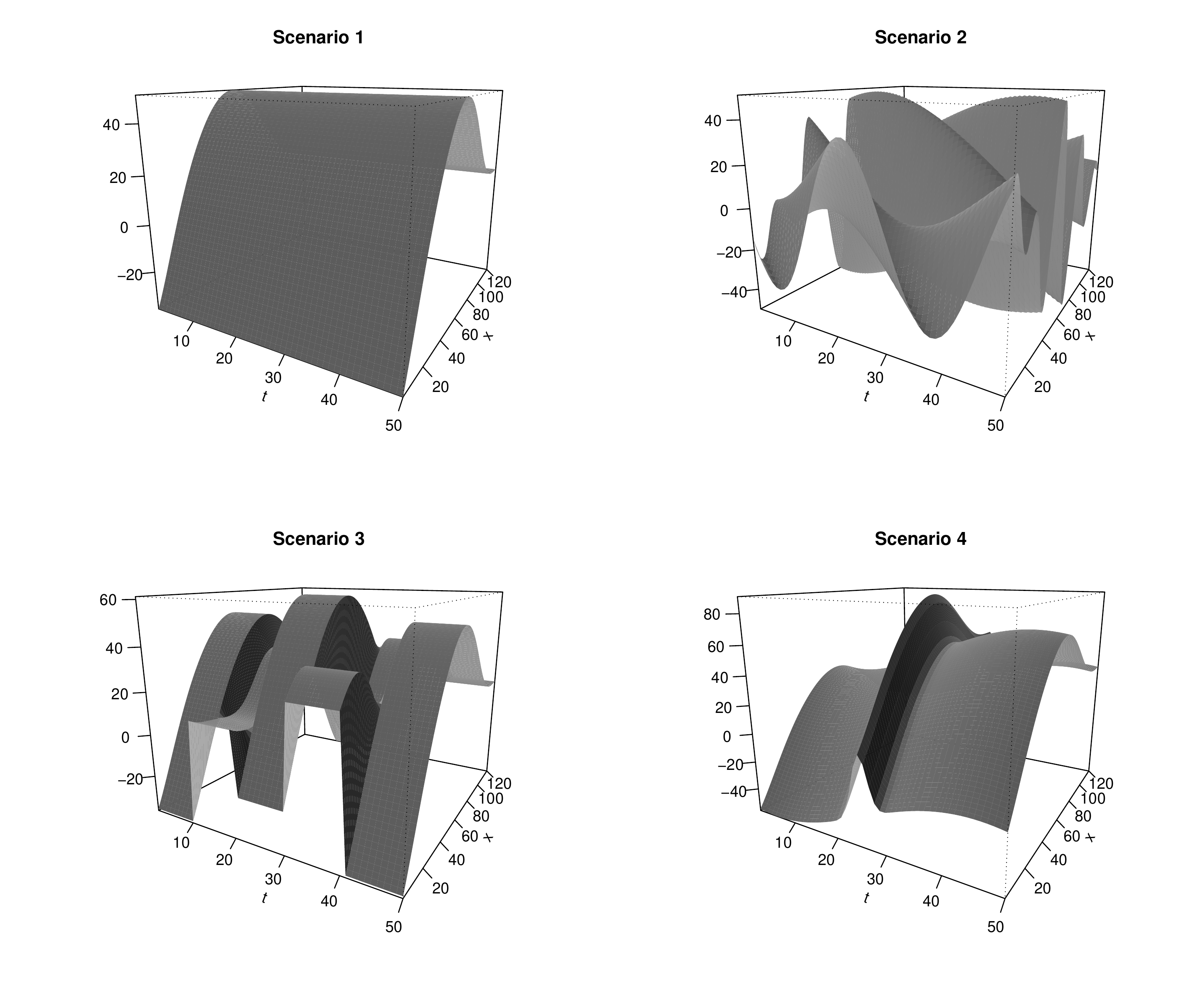}
  \end{center}
  \caption{Each surface represents a three-dimensional plot of the true trend.\label{fig:data_3d}}
\end{figure}

We investigated the performance of the proposed methods together with existing ones through simulation studies. 
For $t=1,...,T \,(=50)$ and the domain $\mathcal{X}=[1,120]$, we adopted the four scenarios of the true trend function:
\begin{align*}
&\text{(1) Constant:}  \ \ \beta_t(x)=f_1(x),\\
&\text{(2) Smooth:} \ \ \beta_t(x)=f_1(x)\sin\frac{t+x}{5},\\ 
&\text{(3) Piecewise constant:} \ \ 
\beta_t(x)=f_1(x)\mathbb{I}_{\{0< t\leq 10\}} + f_2(x)\mathbb{I}_{\{10< t\leq 20\}} + f_3(x)\mathbb{I}_{\{20< t\leq 30\}} \\
&+ f_4(x)\mathbb{I}_{\{30< t\leq 40\}} + f_5(x)\mathbb{I}_{\{40< t\leq 50\}} ,\\
&\text{(4) Varying\ smoothness:} \ \ 
\beta_t(x)=f_1(x)+20\left\{\sin\left(\frac{4t}{n}-2\right)+2\exp\left(-30\left(\frac{4t}{n}-2\right)^2\right)\right\},
\end{align*}
where $f_1,...,f_5$ are sample paths of the Gaussian process associated to RBF kernel $k(x_1,x_2) = \theta^2 \exp(-\|x_1-x_2\|^2/(2\theta^2))$ with a hyper-parameter $\theta$. We set $\theta= 30, 20, 35, 25, 30$ respectively for $f_1,f_2,...,f_5$.
The observed functional data were generated by adding $N(0,\sigma^2)$ noise at equally spaced $H=120$ points of $x$, namely, $x\in \{1,2,\ldots,H\}$.
The trends of the functions under the four scenarios are shown in Figure \ref{fig:data_3d}.

In scenario (1), we examine the abilities of the methods to find the horizontal line in the presence of noise. In scenario (2), we investigate whether the adaptive methods extract the continuous curve from the noisy data. Scenario (3) unearths the capability of the methods to spot the sharp changes, the points of discontinuities, between intermittent straight horizontal lines. In scenario (4), we test the abilities to catch the trend when the smoothness of the process varies significantly due to a sharp peak in the middle as a function of $t$.

\subsection{\textit{Functional trend filtering}}
\label{sec:sim:ftf}
For the simulated data, we apply the following three methods:
\begin{itemize}
\item[-]
{\bf FTF}: Functional trend filtering with $k\in \{0, 1, 2\}$.

\item[-]
{\bf FHP}: Functional HP filter with $k\in \{0, 1, 2\}$.

\item[-]
{\bf FPC}: The standard functional principle component method using R package ``fda.usc".
\end{itemize}

\begin{figure}[]
  \begin{center}
  \includegraphics[width=16cm]{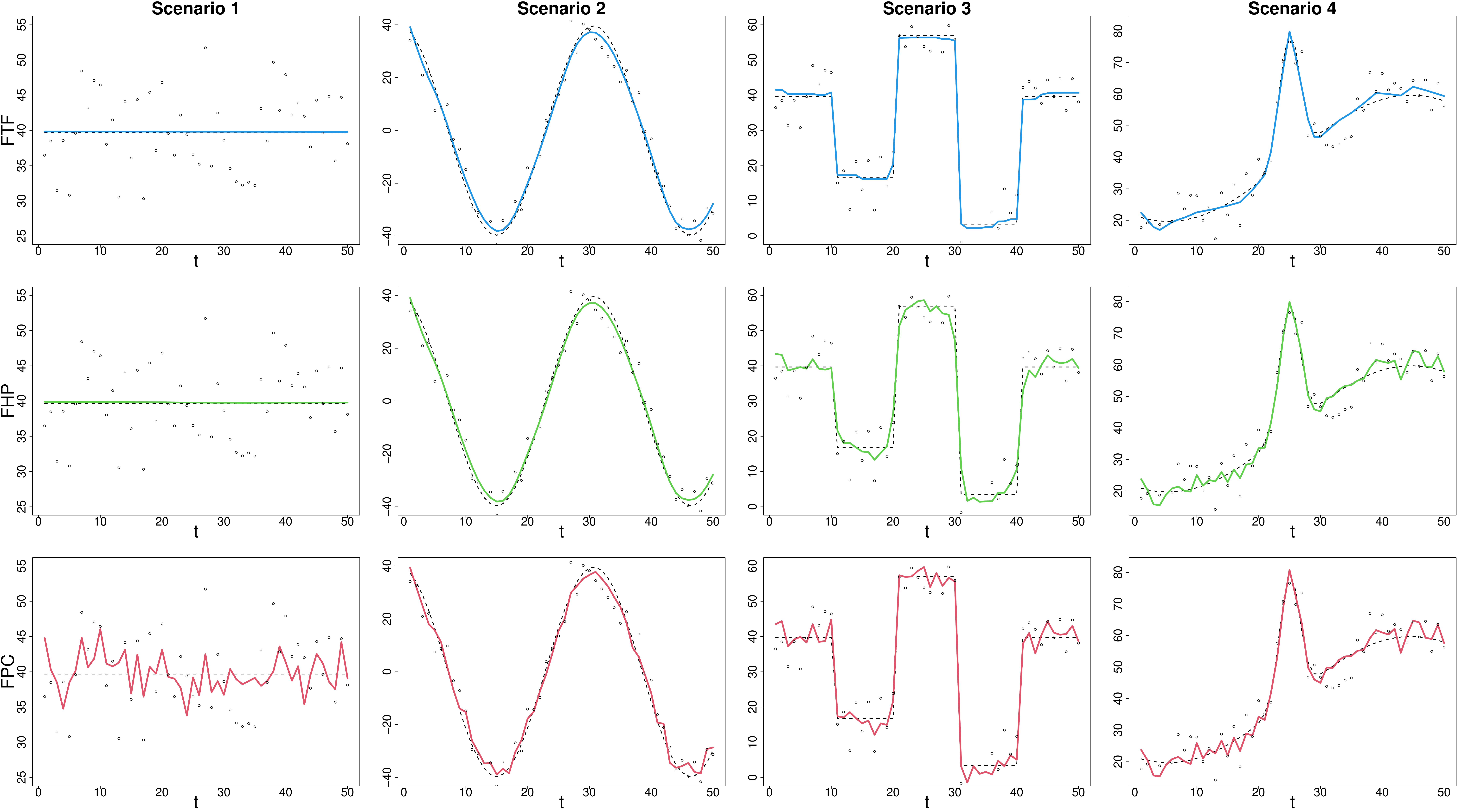}
  \end{center}
  \caption{Plots show data points and the fitted results of the three methods, FPC, FTF and FHP at $x=40$ under four scenarios with noise level $\sigma=5$. The order $k$ of the methods is $0,2,0$ and $1$ for each scenario.\label{fig:sim-fit} }
\end{figure}

\begin{table}
\caption{MSE of functional trend filtering (FTF), functional HP filtering (FHP) and functional principle component analysis (FPC) under four scenarios with $L=5$ and $\sigma\in \{3,5,7\}$.\label{tab:sim} } 
\centering
\medskip
\begin{tabular}{ccccccccccccc} 
\hline
& & \multicolumn{4}{c}{Scenario} \\
$\sigma$ & method &  & (1) & (2) & (3) & (4)\\ 
\hline
&FTF ($k=0$)&  & 0.190 & 1.801 & 1.139 & 1.098\\ 
&FTF ($k=1$)&  & 0.316 & 1.050 & 1.721 & 1.095\\   
&FTF ($k=2$)&  & 0.426 & 0.987 & 1.718 & 1.261\\  
3&FHP ($k=0$)& & 0.177 & 1.803 & 1.919 & 2.099\\  
&FHP ($k=1$)&  & 0.255 & 1.054 & 1.911 & 2.073\\  
&FHP ($k=2$)&  & 0.350 & 0.998 & 1.912 & 1.996\\   
&FPC        &  & 2.176 & 1.959 & 1.923 & 2.112\\  
\hline
&FTF ($k=0$)&  & 0.490 & 3.917 & 2.964 & 3.291 \\ 
&FTF ($k=1$)&  & 0.668 & 2.787 & 3.900 & 2.399\\  
&FTF ($k=2$)&  & 0.891 & 2.307 & 4.116 & 2.839\\  
5&FHP ($k=0$)& & 0.491 & 3.873 & 5.165 & 5.712 \\ 
&FHP ($k=1$)&  & 0.710 & 2.589 & 5.160 & 5.140\\  
&FHP ($k=2$)&  & 0.940 & 2.064 & 4.876 & 4.132\\  
&FPC        &  & 6.045 & 5.510 & 5.220 & 5.750 \\ 
\hline
&FTF ($k=0$)&  & 0.973 & 6.344 & 4.841 & 5.087\\ 
&FTF ($k=1$)&  & 1.671 & 4.499 & 6.776 & 4.485\\   
&FTF ($k=2$)&  & 2.153 & 4.338 & 7.927 & 5.147\\  
7&FHP ($k=0$)& & 0.964 & 6.072 & 9.031 & 9.632\\  
&FHP ($k=1$)&  & 1.393 & 4.266 & 8.973 & 7.431\\  
&FHP ($k=2$)&  & 1.907 & 4.071 & 9.445 & 6.685\\  
&FPC        &  & 11.848 & 10.403 & 9.662 & 11.209\\  
\hline
\end{tabular}
\end{table}

Note that we used the estimated principle component functions by FPC as orthonormal functions for FTF and FHP with $L=5$ (the number of principle functions) to allow comparison independent of basis functions. By $10$-fold cross-validation, we select the tuning parameter $\lambda$ from the space $[10^{-3},10^{3}]$ by checking $60$ points equally spaced on a logarithmic scale in all scenarios.

The estimated trend functions at $x=40$ are presented in Figure \ref{fig:sim-fit}.
Based on 150 times repeated simulation, we also report the mean squared error (MSE):\begin{align*}
    \mathrm{MSE}=\frac{1}{TH}\sum_{t=1}^T\sum_{x=1}^H(\hat{\beta}_t(x)-\beta_t(x))^2,
\end{align*}
in Table \ref{tab:sim}, where  $\hat{\beta}_t(x)$ is the estimated function.
Overall, the proposed FTF tended to perform better than the other methods.
Further, we can see from Figure \ref{fig:sim-fit} that FPC provided under-smoothed trend estimate compared with FTF and FHP, which is related to the overall performance in terms of MSE reported in Table \ref{tab:sim}.

Interestingly, the performance of FTF and FHP were quite different, although the only methodological difference is whether $L^2$-norm or squared $L^2$-norm is adopted in the penalty.
For example, in scenario 3, the performance of FHP was almost the same as that of FPC while FTF provided better results. This is attributed to the fact that FHP does not produce sparsity. Regarding the performance of FTF depending on $k$, it is observed that FTF with $k=0$ provided the most accurate results in scenario 3 since the true trend admits a piecewise constant structure that FTF with $k=0$ is considered to work well.
In the other scenarios, however, the piecewise constant structure seems rather limited, and the performance of FTF with $k=1,2$ is more appealing. 
It is worth noting that FTF with $k=1$ performed outstandingly well in scenario 4.
Around the peak of scenario 4, the smoothness of the trend changes abruptly.
The change in smoothness is nearly equal to the change in the amount of difference.
Hence, the sharp peak of scenario 4 is the point where the supremacy of FTF exists.
By contrast, in Scenario 1 and Scenario 2, FTF is slightly inferior to FHP.
One possible reason is that FTF is a numerical solution obtained by iterative approximation while FHP is an analytical solution.

\subsection{\textit{Sparse functional trend filtering}}
\label{sec:sim:sftf}

\begin{table}[]
\caption{MSEs and the corresponding number of basis functions (in parentheses) of sparse functional trend filtering (SFTF) and dynamic functional principle component analysis (DFPC) under four scenarios with $\sigma\in \{3,5,7\}$.\label{tab:sim-sftf} }
\centering
\medskip
\begin{tabular}{ccccccccccccc} 
\hline
& & \multicolumn{4}{c}{Scenario} \\
$\sigma$ & method &  & (1) & (2) & (3) & (4)\\ 
\hline
&SFTF ($k=0$)& & 0.177 & 0.696 & 0.955 & 0.450\\ 
&            & & (0.00)& (2.00)& (3.14)& (1.00) \\ 
3&SFTF ($k=1$)&& 0.177 & 0.644 & 1.031 & 0.445\\  
&            & & (0.00)& (2.00)& (3.20)& (1.00) \\ 
&SFTF ($k=2$)& & 0.177 & 0.666 & 1.041 & 0.440\\  
&            & & (0.87)& (2.00)& (3.26)& (1.00) \\ 
&DFPC       & & 1.510 & 1.576 & 1.510 & 1.511\\  
\hline
&SFTF ($k=0$)& & 0.490 & 1.902 & 2.245 & 1.221  \\ 
&            & & (3.15)& (2.76)& (4.99)& (1.49) \\ 
5&SFTF ($k=1$)&& 0.491 & 1.700 & 2.551 & 1.213  \\  
&            & & (0.84)& (2.83)& (3.51)& (1.49) \\ 
&SFTF ($k=2$)& & 0.491 & 1.745 & 2.631 & 1.196  \\  
&            & & (2.67)& (2.91)& (3.53)& (1.47) \\ 
&DFPC & & 4.187 & 4.254 & 4.188 & 4.189  \\ 
\hline
&SFTF ($k=0$)& & 0.961 & 3.670 & 4.350 & 2.370\\ 
&            & & (5.88)& (2.57)& (6.43)& (1.51) \\ 
7&SFTF ($k=1$)&& 0.961 & 3.270 & 4.773 & 2.345\\ 
&            & & (2.06)& (3.55)& (4.27)& (1.51) \\ 
&SFTF ($k=2$)& & 0.962 & 3.275 & 4.996 & 2.321 \\  
&            & & (1.62)& (3.27)& (4.46)& (1.51) \\ 
&DFPC        & & 8.204 & 8.272 & 8.205 & 8.206\\  
\hline
\end{tabular}
\end{table}

The above simulation showed that FTF accurately estimated the trend even with sudden changes.
However, we need to choose the appropriate number of basis functions.
Hence, we applied SFTF, introduced in Section \ref{sec:NoB}, and investigated whether the number of basis functions could be selected. Specifically, we set $L=10$ first, and then applied SFTF and cut off unnecessary basis functions. We implemented 150 simulations and calculated MSE of SFTF and mean of the number of basis functions. We searched for the optimal values of $(\lambda,\psi)$ from the space $[10^{-3},10^{3}]\times [10^{-1},10^{1}]$ by checking $60\times20$ points equally spaced on a logarithmic scale in all scenarios.
As a competitor, we apply an advanced method, dynamic functional principal component analysis (DFPC), which incorporate serial dependence.
The R package ``freqdom.fda" does not mention anything about parameter selection. Then, we set the parameters to minimize the MSE. Namely, we compare SFTF to this DFPC with oracle parameters.

Table \ref{tab:sim-sftf} presents the MSE of DFPC and that of SFTF.
We chose parameters that favored DFPC, but SFTF dominated it and hence the superiority of SFTF is solidified.
Moreover, the number of basis functions whose coefficients were not set to zero by the SFTF was fairly smaller than 10. In particular, in scenario 2 and 4, the number of selected components was far lower than 5, indicating that many unnecessary components were used for simple FTF. This implies that the accuracy of SFTF was substantially improved by that amount.
Hence choosing the number of basis functions by excluding redundant ones plays a critical role in increasing the accuracy.

\section{Applications}
\label{sec:app}

\subsection{\textit{Australian fertility rates}}
\label{sec:aus}
\begin{figure}[htb]
  \begin{center}
  \includegraphics[width=14cm]{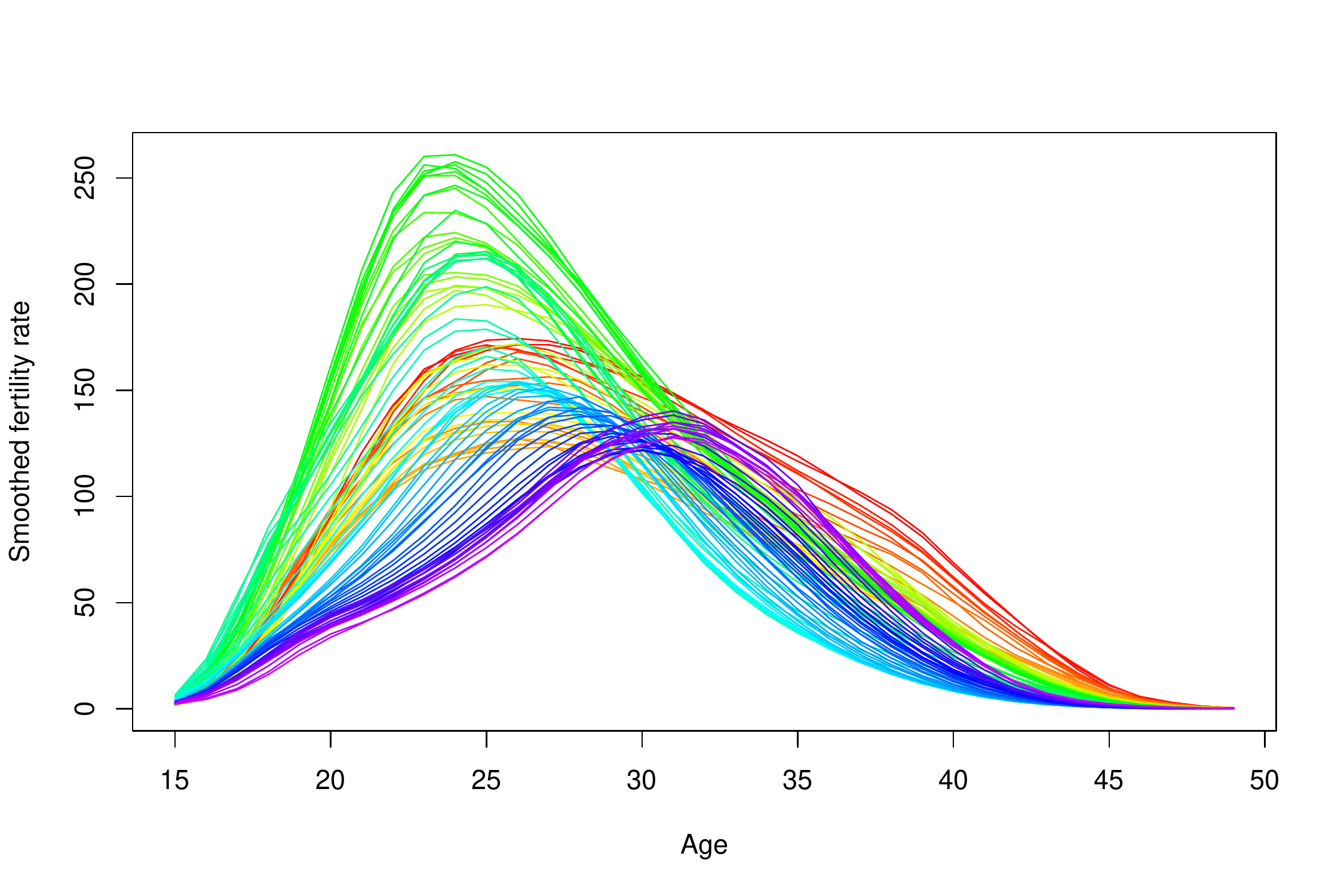}
  \end{center}
\caption{Age-specific Australian fertility rates curves for ages 15 to 49 observed from 1921 to 2015 (in the same order as the color in a rainbow).\label{fig:rainbow}}
\end{figure}

Fertility rates in Australia have been declining seriously as in other developed countries. We examined the data "Australiasmoothfertility", which is available from R package ``rainbow". 
The original data, obtained from the Australian Bureau of Statistics, describes the age-specific number of live births per 1000 females of ages $15,16,...,49$ from 1921 to 2015. 
The data is functional data and each function represents the age-specific number between $15$ and $49$ in a year. Fig \ref{fig:rainbow} shows the curves with rainbow colors. The colors indicate that the oldest curve is red, the newest curve is purple and the others are colored in the same order as a rainbow.

Here we applied SFTF to the dataset and set $L=10$ first.
We selected tuning parameters $(\lambda,\psi)$ from the space $[10^{-2},10^{2}]\times [10^{-1},10^{0}]$ by checking $40\times10$ points equally spaced on a logarithmic scale in all scenarios.
\begin{figure*}[h]
  \begin{center}
  \includegraphics[width=15.5cm]{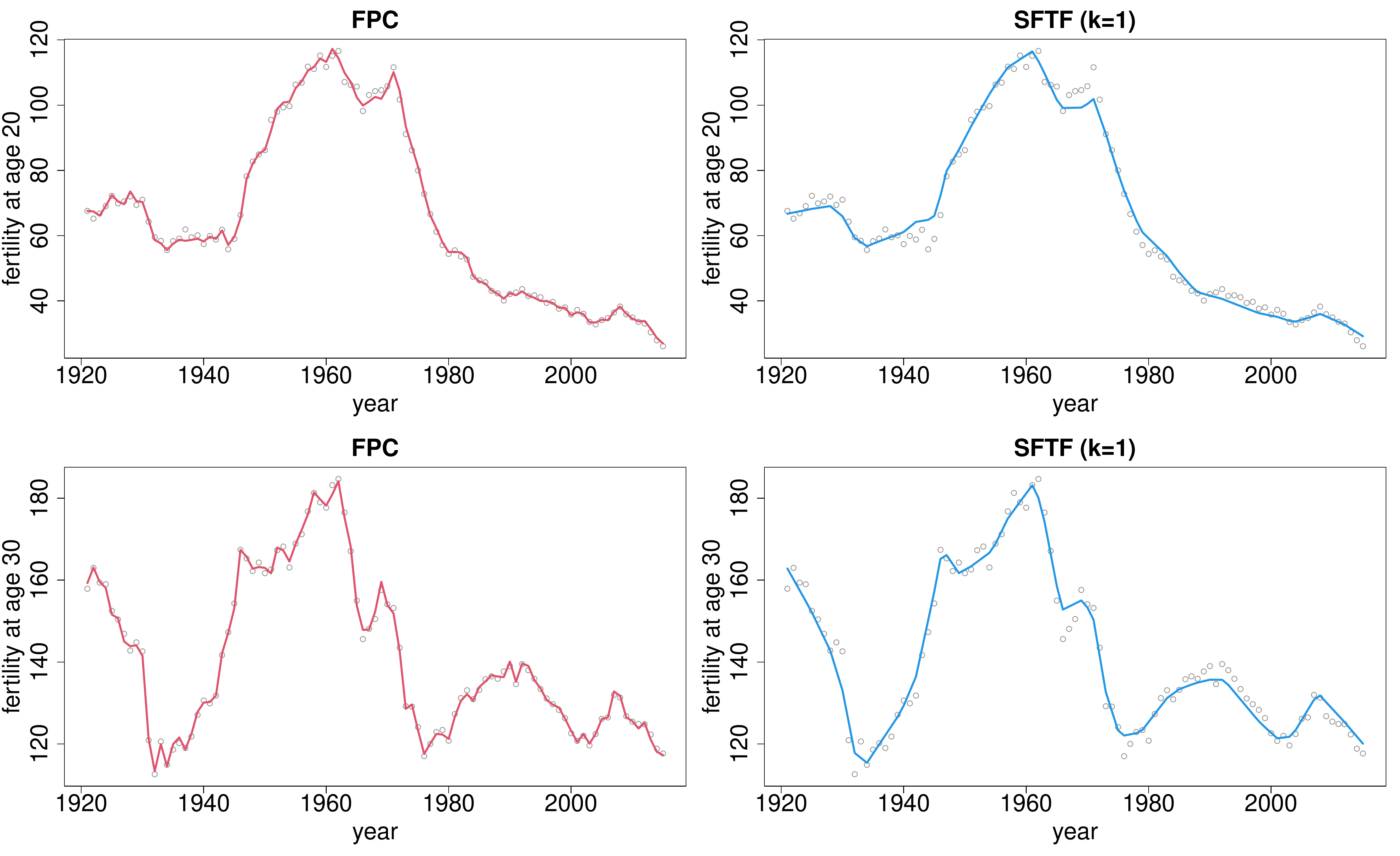}
  \end{center}
  \caption{At age $20$ and $30$, the number of births per $1000$ females (round points), trends fitted by FPC (red lines in left column), and trends fitted by 1st order SFTF (blue lines in right column).
\label{fig:fertility}}
\end{figure*}

\begin{figure}[htb]
  \begin{center}
  \includegraphics[width=15.5cm]{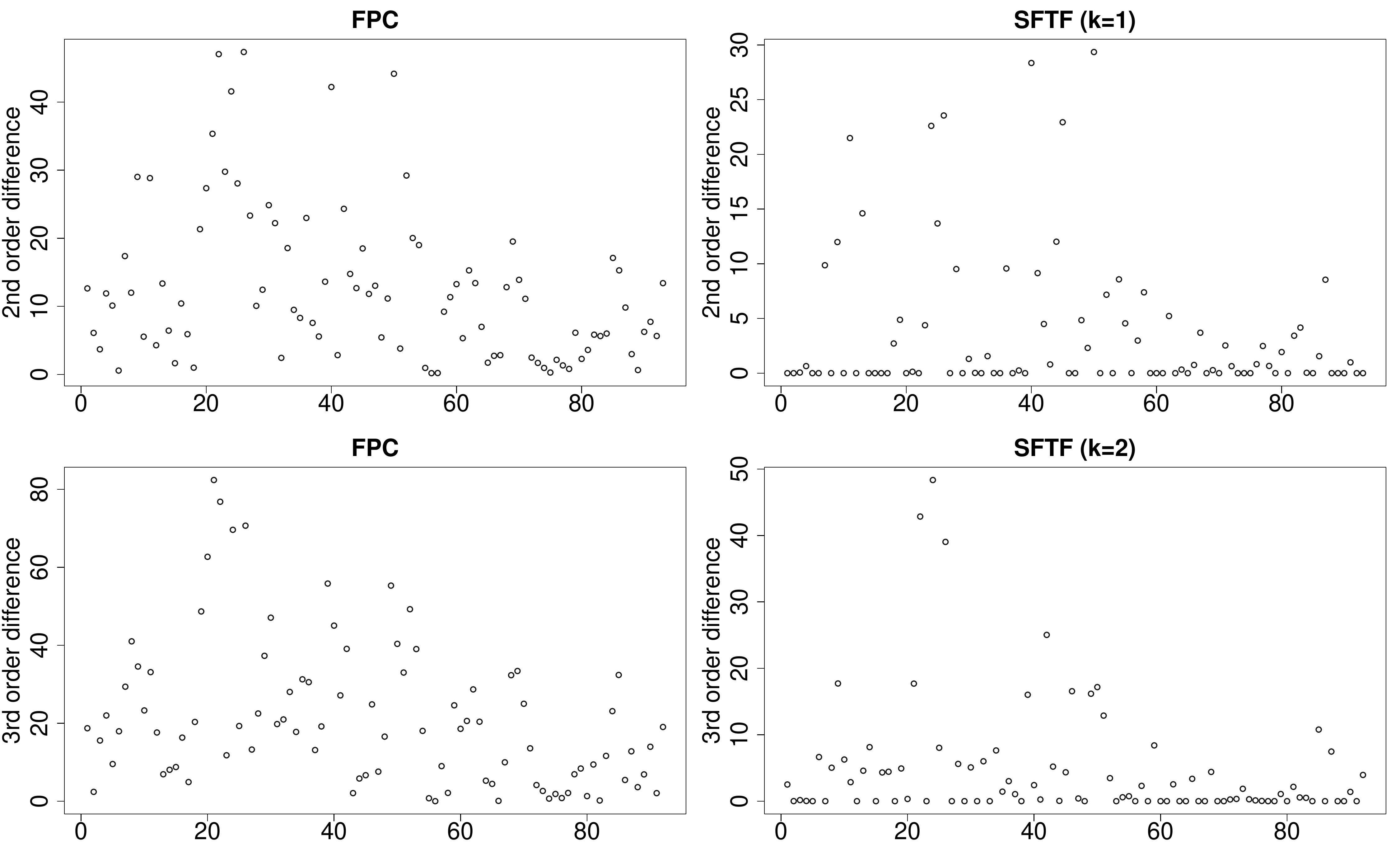}
  \end{center}
  \caption{Top left: absolute values of 1st order differences in scores of PC1. Top right: absolute values of 1st order differences in the trend filtered scores. Bottom left: absolute values of 2nd order differences in the scores. Top right: absolute values of 2nd order differences in the trend filtered scores.}\label{fig:asf_diff}
\end{figure}

Figure \ref{fig:fertility} shows the number of births per 1000 females of ages 20 and 30 in all years and curves fitted by FPC and SFTF. Figure \ref{fig:asf_diff} displays absolute values of 1st order differences and 2nd order differences in scores of the first principal component (PC1) between the years and their trend filtered versions. First, compared with FPC, the ability of SFTF to serve as a smoother is confirmed from Figure \ref{fig:fertility}.
It eliminates small noises, but retains the significant change points.

Next, in common between age 20 and 30 in Figure \ref{fig:fertility}, we find abrupt changes in 1961 and 1972. After World War I\hspace{-.01em}I, the fertility rate had increased until 1961, although the first oral contraceptive pill was released in Australia in 1961. Furthermore, in 1972, the prime minister of Australia at that time abolished the 27.5 percent luxury tax on all contraceptives \citep{mclennan1998}. It increased the use of the pills especially among young people and the trends are reflected as the sharp change points in plots in Figure \ref{fig:fertility}. 
Moreover, from the upper right plot in Figure \ref{fig:asf_diff}, the structure in the sense of 2nd order difference is considered to change at 40th and 50th points; namely, large structural changes occur from 1960 to 1962 and from 1970 to 1972. The lower left plot suggests, in terms of 2nd order difference, the structure changes at 24th point (i.e. around 1945), implying that the trend of the fertility rate changed after the end of World War I\hspace{-.01em}I. Owing to the sparsity in differences in trend, we easily detect those underlying events. In addition, since the detected points from the plots tend to be overlapped between $k=1$ and $2$ in Figure \ref{fig:asf_diff}, trend filtering would be able to stably extract the turning points regardless of the order $k$.
By contrast, we hardly find the structural properties of data from the original scores of the principal component.
Therefore, the result proves the ability of trend filtering to catch sharp changes.

\subsection{\textit{The number of COVID-19 cases in Japanese prefectures}}

Infection with the novel COVID-19 has been spreading since 2020 and has brought about many deaths worldwide. Thus analyzing the situation becomes increasingly important. For instance, \cite{tang2020functional} exploited some functional time series methods to analyze the COVID-19 data in the US. 
In this study, we investigate the number of people infected by COVID-19 by prefecture in Japan, which is available at \url{https://www3.nhk.or.jp/news/special/coronavirus/data-widget/}, and we scale the number by population of each prefecture available at \url{https://www.stat.go.jp/data/nihon/02.html}.
Each prefecture is treated as a vertex on a graph, and when the prefectures are adjacent to each other, the connectivity of the graph is considered. We handled the number of infected people per million in each prefecture from January 16, 2020 to March 9, 2021, and regarded them as functional data after smoothing.

\begin{figure*}[tb]
  \begin{center}
  \includegraphics[width=12.6cm]{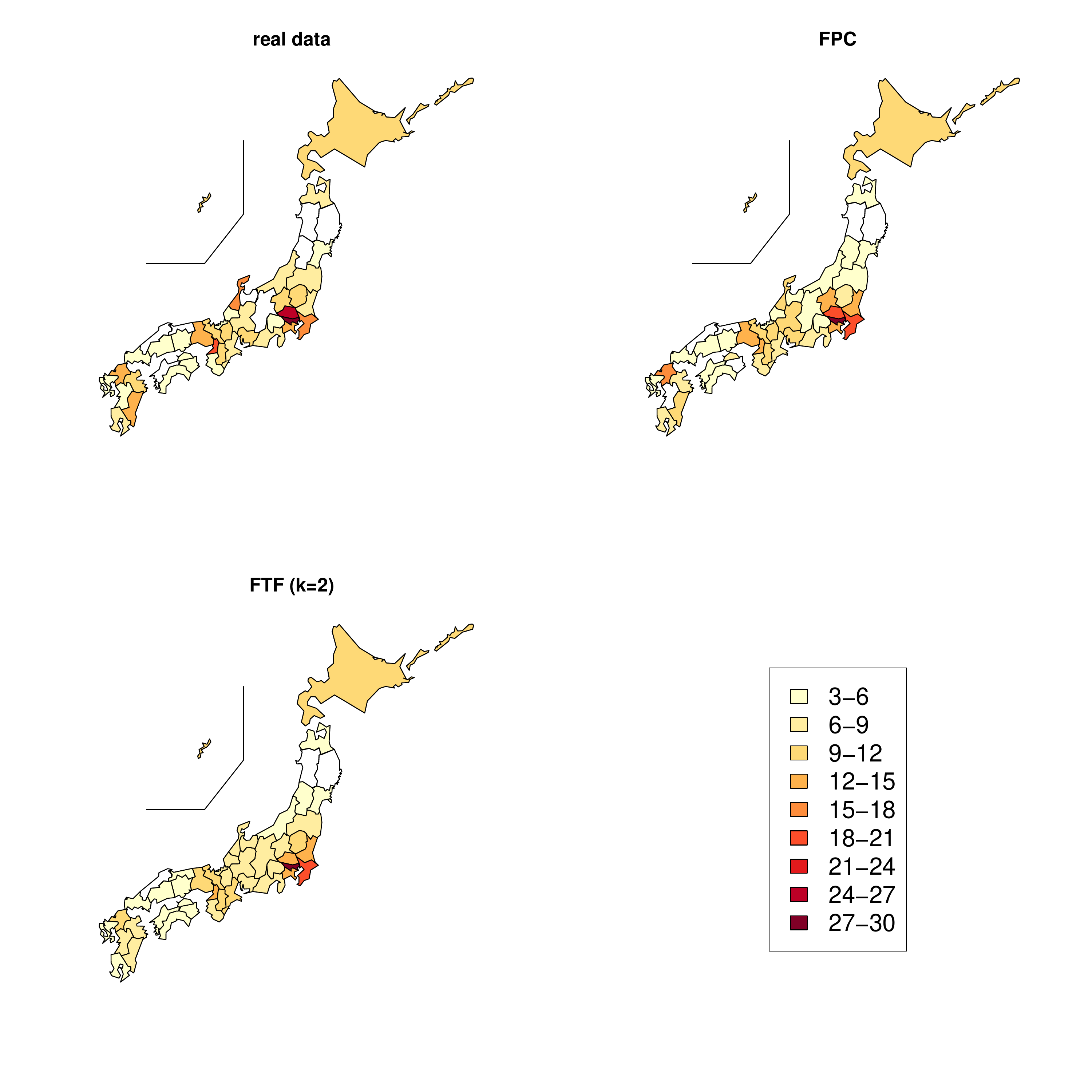}
  \end{center}
  \caption{Top left: The observed number of infected people by prefecture on day 395. Top right: The number of infected people by prefecture on day 395, smoothed by FPC.  Bottom left: The number of infected people by prefecture on day 395, smoothed by 2nd order FTF.}\label{fig:covid}
\end{figure*}

The observed data on 395th day are shown in the upper left panel in Figure \ref{fig:covid}. We plot the data fitted by FPC in the upper right panel and FTF with $k=2$ in the lower left panel, where the value of $\lambda$ was selected as the argument of the minimum MSE from $[10^{-3},10^3]$. We also applied FTF with $k=1$, but the result is almost the same, thereby we do not display it here.

For a qualitative visual analysis, although FTF was smooth trend better than FPC, whose result was still jagged, as we have expected, trend filtering was more effective in that it spotted an outstanding (dark colored) prefecture, Tokyo. 
Evidently, FTF is able to localize its estimates around strong inhomogeneous spikes, which implies that it is able to detect the event or spot of interest.

\section{Discussion}\label{sec:conc}
In this paper, we proposed a functional version of the locally adaptive smoothing technique known as trend filtering for smoothing functional time series and spatial data. 
The need to consider group lasso + fused lasso like penalty allows for a trivial extension of the scalar version, but we developed an efficient optimization algorithm to obtain trend estimation and discussed the choice of tuning parameter. 
Through simulation and empirical studies, we demonstrated the superiority of the proposed method to existing methods. 

Moreover, in time series data, we can select the number of basis functions by adding a penalty.
The reduction of unnecessary basis functions denoises the functions themselves, whereas trend filtering is smoother with respect to time direction.
As a result, the performance of the simulation is also improved, showing that choosing the number of basis functions is better than just taking more basis functions.
On the whole, penalty is the key to the methods we developed.

The optimization problem for computing the proposed method can be regarded as generalization and combination of grouped and fused lasso estimation, thereby it would be interesting to apply the proposed optimization techniques to other statistical problems, for example, regression analysis with complicated sparsity-inducing penalty functions.

\section*{Acknowledgment}
This research is partially supported by Japan Society for Promotion of Science (KAKENHI) grant numbers 18H03628 and 21H00699.

\vspace{1cm}
\newpage
\appendix
\begin{center}
{\large {\bf  Appendices}}
\end{center}

\section*{Appendix 1: Derivation of Algorithm 1}
We here provide the detailed derivation of each step in Algorithm 1. 

\begin{itemize}

\item[-]
(Update of $\bm{b}_{\ell}$)\ \ 
For $\ell=1,\ldots,L$, given $u_{t\ell}$ and $\bm{a}_t$, $\bm{b}_{\ell}$ is updated by using the minimizer of 
\begin{align*}
\frac12\sum_{\ell}\|\bm{z}_{\ell}-\bm{b}_{\ell}\|_2^2  +  \sum_t\sum_{\ell} u_{t\ell}(\bm{e}_t^a\Delta\bm{b}_{\ell}-\bm{e}_{\ell}^b\bm{a}_t)+\frac{\rho}{2}\sum_t\sum_{\ell}(\bm{e}_t^a\Delta\bm{b}_{\ell}-\bm{e}_{\ell}^b\bm{a}_t)^2,
\end{align*}
which is a quadratic function of $\bm{b}_{\ell}$.
Since its derivative with respect to $\bm{b}_{\ell}$ is given by 
\begin{align*}
(I+\rho \Delta^{\top}\Delta)\bm{b}_{\ell} - \bm{z}_{\ell}  +  \sum_{t}u_{t\ell}(\bm{e}_t^a\Delta)^{\top}-\rho\sum_{t}(\bm{e}_t^a\Delta)^{\top}\bm{e}_{\ell}^b\bm{a}_t,
\end{align*}
the minimizer can be obtained as 
\begin{align*}
\bm{b}_{\ell}
\leftarrow
(I+\rho \Delta^{\top}\Delta)^{-1}
\left\{\bm{x}_{\ell}  -  \sum_{t}u_{t\ell}(\bm{e}_t^a\Delta)^{\top}  +  
\rho\sum_{t}(\bm{e}_t^a\Delta)^{\top}\bm{e}_{\ell}^b\bm{a}_t \right\}.
\end{align*}

\item[-]
(Update of $\bm{a}_{t}$) \ \ 
For $t=1,...,T-k-1$, given $u_{t\ell}$ and $\bm{b}_{\ell}$, $\bm{a}_{t}$ is updated as the minimizer of 
\begin{align*}
\label{a_t}
\lambda \sum_{t=1}^{T-k-1}\|\bm{a}_t\|_2 
+  \frac{\rho}{2}\sum_t\sum_{\ell}\left(\bm{e}_t^a\Delta\bm{b}_l-\bm{e}_{\ell}^b\bm{a}_t+\frac{u_{t\ell}}{\rho}\right)^2
\end{align*}
Because the objective function is non differentiable due to the presence of the $\|\bm{a}_t\|_2$, we deal with the problem by a proximal method. We denote the first term (non-smooth part) by $f_{nsm}$ and the second part (smooth part) by $f_{sm}$. Since $f_{sm}$ is convex and $\nabla_{\bm{a_t}}f_{sm}$ is Lipschitz continuous with constant 1, FISTA (fast iterative shrinkage-thresholding algorithm), first presented by \cite{beck2009fast}, can be applied. Remarking that, in general, the proximity operator of the $\ell_2$ norm ($\lambda \|\cdot\|_2:\mathbb{R}^d\rightarrow \mathbb{R},\,d\in\mathbb{N}$), known as soft thresholding operator, is $S_{\lambda}(\bm{s})=\max(0, 1-\lambda/\|\bm{s}\|_2)\bm{s}$ for $\bm{s}\in \mathbb{R}^d$, we get the updating step given in Algorithm 1.

\item[-]
(Update of $u_{t\ell}$) \ \ 
For $\ell=1,\ldots,L$ and $t=1,...,T-k-1$, given $\bm{b}_{\ell}$, $\bm{a}_{t}$ and $u_{t\ell}^{\ast}$ (current value of $u_{t\ell}$), $u_{t\ell}$ is updated as $u_{t\ell}\leftarrow u_{t\ell}^{\ast}+\rho(\bm{e}_t^a\Delta\bm{b}_{\ell}-\bm{e}_{\ell}^b\bm{a}_t)$.

\end{itemize}

\section*{Appendix 2: Algorithm of sparse functional trend filtering}
We introduce the computational algorithm of sparse functional trend filtering. The difference from Algorithm 1 is the update part of $\{\bm{b}_{\ell}\}$. We derived the way to update $\{\bm{b}_{\ell}\}$ by using FISTA as the update step of $\{\bm{a}_t\}$ in Algorithm 1.

\bibliographystyle{chicago}

\bibliography{ref}

\end{document}